# The elastic collision of two H-like atoms including the exotic Ps and Mu

Hasi Ray[1,2,3]

[1] Study Center, S-1/407/6, B. P. Township, Kolkata 700094, India
[2] Department of Physics, New Alipore College, Kolkata 700053, India
[3] Science Department, National Institute of TTT and Research, Kolkata 700106, India

Email:  hasi_ray@yahoo.com   and   hasi.ray1@gmail.com

**Abstract**: The elastic collision between two H-like atoms utilizing an ab-initio static-exchange model (SEM) in the center of mass (CM) frame considering the system as a four-body Coulomb problem where all the Coulomb interaction terms in the direct and exchange channels are treated exactly, is studied thoroughly. A coupled-channel methodology in momentum space is used to solve Lippman-Schwinger equation following the integral approach. The new SEM code [1-5] in which the Born-Oppenheimer (BO) scattering amplitude acts as input to derive the SEM amplitude using partial wave analysis, is utilized to study the s-, p-, d- wave elastic phase shifts and the corresponding partial cross sections. An augmented-Born approximation is used to include the contribution of higher partial waves more accurately to determine the total/integrated elastic cross sections. The effective range theory is used to determine the scattering lengths and effective ranges in the s-wave elastic scattering. The systems studied are Ps-Ps, Ps-Mu, Ps-H, Ps-D, Ps-T, Mu-Mu, Mu-H, Mu-D, Mu-T, H-H, H-D, H-T, D-D, D-T, T-T. The SEM includes the non-adiabatic short-range effects due to exchange. The MSEM code [1] is used to study the effect of the long-range van der Waals interaction due to induced dipole polarizabilities of the atoms in H(1s)-H(1s) elastic collision. The dependence of scattering length on the reduced mass of the system and the dependence of scattering length on the strength of long-range van der Waals interaction that varies with the minimum interatomic distance are observed [3].

1. Introduction

   The elastic collision between two hydrogen-like atoms is of fundamental interests because the wavefunctions of both the atoms are perfectly known. It is a four-body Coulomb problem [1-6]. Recently Ray [1-5] evaluated exactly all the Coulomb interaction terms in such two-atomic system for both the direct and exchange channels and developed a new FORTRAN code. The static-exchange model (SEM) was used to calculate the s-, p-, d- wave elastic phase shifts and corresponding partial cross sections; the effective range theory was used to evaluate the scattering lengths and effective ranges for the s-wave. The augmented-Born approximation was used to include the contribution of higher partial waves as accurately as possible to calculate the total/integrated elastic cross section. All the atoms are considered at their ground states. Different isotopes used are Positronium (Ps), Muonium (Mu), Hydrogen (H), Deuterium (D), Tritium (T). In the systems like Ps-Mu, Ps-H, Ps-D, Ps-T and Ps-Ps, due to light mass of projectile the s-wave contribution dominates. The reduced masses of these systems are very close to 2. In Ps-Ps system the reduced mass is 1. When both the atoms are heavier e.g. Mu-Mu, Mu-H, Mu-D, Mu-T, H-H, H-D, H-T, D-D, D-T, T-T, the s-wave contributions are important but not dominating. Here the reduced masses are much higher and very different from each other. In two-atomic collision at low and cold energies, the non-adiabatic interaction due to exchange of electrons and the long-range van der Waals interaction due to induced dipole-polarizabilities of the atoms are important. Recently Ray [1] introduced a modified static exchange model (MSEM) to include both the interactions. The SEM adapted by Ray [1-7] included the non-adiabatic effects due to exchange of electrons between the atoms. In the MSEM [1] code, the SEM potential is modified by including the effect of van der Waals interaction between the atoms. Here the strength of van der Waals interaction varies with minimum interatomic distance ($R_0$). The reduced masses ($\mu$) of the systems in atomic units are respectively : (i) $\mu = 103.9$ for Mu and Mu, (ii) $\mu = 186.7$ for Mu and H, (iii) $\mu = 196.7$ for Mu and D, (iv) $\mu = 200.2$ for Mu and T, (v) $\mu = 918.5$ for H and H, (vi) $\mu = 1224.5$ for H and D, (vii) $\mu = 1377.6$ for H and T, (viii) $\mu = 1836.5$ for D and D, (ix) $\mu = 2203.7$ for D and T, (x) $\mu = 2754.5$ for T and T.

   The paper is planned to provide using the SEM code: (i) the s-, p-, d- wave elastic phase shifts, the corresponding partial cross sections and the total/integrated elastic cross sections for collision between two H(1s) atoms; (ii) the s-, p-, d- wave elastic phase shifts, the corresponding partial cross sections and the total/integrated elastic cross sections and quenching cross sections for collision between two Ps(1s) atoms; (iii) The variation of singlet and triplet scattering lengths with reduced masses for systems involving Ps e.g. Ps-Ps, Ps-Mu, Ps-H, Ps-D and Ps-T; (iv) The variation of triplet scattering lengths with reduced masses when both the atoms are heavier e.g. Mu-Mu, Mu-H, Mu-D, Mu-T, H-H, H-D, H-T, D-D, D-T, T-T. Next (v) the MSEM code is used to study the variation of triplet scattering lengths in H(1s)-H(1s) elastic scattering with the variation of the strength of interatomic potential e.g. the van der Waals interaction, varying the minimum interatomic distance $R_0$.

2. Theory

   The detailed theories are available in references [1-7] and for convenience it is briefly described here. The initial and final state wavefunctions are defined as

$$\psi_i = e^{i\vec{k}_i \cdot \vec{R}'} \phi_{1s}^A(\vec{r}_1)\phi_{1s}^B(\vec{r}_2) \qquad (2.1a)$$

$$\psi_f = (1 \pm P_{12})e^{i\vec{k}_f \cdot \vec{R}_f} \phi_{1s}^A(\vec{r}_1)\phi_{1s}^B(\vec{r}_2) \qquad (2.1b)$$

if $A$ and $B$ are two atoms, $\vec{k}_i$ and $\vec{k}_f$ represent the initial and final momenta of the projectile. In elastic scattering $|\vec{k}_i| = |\vec{k}_f|$; so only the direction of the final momentum $\vec{k}_f$ changes. Here $\phi_{1s}^A(\vec{r}_1)$ and $\phi_{1s}^B(\vec{r}_2)$ are the ground state wave functions of the two atoms and $P_{12}$ is the exchange (or antisymmetry) operator.

Projecting different states on the Schrodinger equation just like the Hartree-Fock variational method one can get the integro-differential equations that can be solved by the method of iteration. Here the Lippman-Schwinger type integral equation in the momentum space formalism [8] rather than using the coordinate space adapted by Fraser in Ref [9] is used. The formally exact Lippman-Schwinger type coupled integral equation for the scattering amplitude in momentum space is given by [8]:

$$f_{n'1s,n1s}^{\pm}(\vec{k}_f,\vec{k}_i) = B_{n'1s,n1s}^{\pm}(\vec{k}_f,\vec{k}_i) - \frac{1}{2\pi^2} \sum_{n''} \int d\vec{k}'' \frac{B_{n'1s,n''1s}^{\pm}(\vec{k}_f,\vec{k}'')f_{n''1s,n1s}^{\pm}(\vec{k}'',\vec{k}_i)}{\vec{k}_{n''1s}^2 - \vec{k}''^2 + i\varepsilon} \qquad (2.2)$$

Here $B^{\pm}$ are the well known Born-Oppenheimer (BO) scattering amplitude [7] in the singlet (+) and triplet (-) channels respectively. In a similar fashion, $f^{\pm}$ indicate the unknown scattering amplitudes for the singlet and triplet states of the two system electrons. Generally the partial wave analysis is used to reduce the three-dimensional integral equation into the one-dimensional form. Here the BO amplitude ($B^{\pm}$) acts as the input to get the SEM amplitude following Eqn. (2.2) and is defined as

$$B_{n'1s,n1s}^{\pm}(\vec{k}_f,\vec{k}_i) = -\frac{\mu}{2\pi} \int d\vec{R}d\vec{r}_1 d\vec{r}_2 \psi_f^*(\vec{R},\vec{r}_1,\vec{r}_2) V(\vec{R},\vec{r}_1,\vec{r}_2) \psi_i(\vec{R},\vec{r}_1,\vec{r}_2) \qquad (2.3)$$

When $\mu$ is the reduced mass of the system and $V(\vec{R},\vec{r}_1,\vec{r}_2)$ is the Coulomb interaction: $V_{Direct}$ is for the direct channel and $V_{Exchange}$ is for the exchange or rearrangement channel.

$$V_{Direct}(\vec{R},\vec{r}_1,\vec{r}_2) = \frac{Z_A Z_B}{R} - \frac{Z_A}{|\vec{R}-\vec{r}_2|} - \frac{Z_B}{|\vec{R}+\vec{r}_1|} + \frac{1}{|\vec{R}+\vec{r}_1-\vec{r}_2|} \qquad (2.4a)$$

$$V_{Exchange}(\vec{R},\vec{r}_1,\vec{r}_2) = \frac{Z_A Z_B}{R} - \frac{Z_A}{|\vec{r}_1|} - \frac{Z_B}{|\vec{r}_2|} + \frac{1}{|\vec{R}+\vec{r}_1-\vec{r}_2|} \qquad (2.4b)$$

The atomic unit (a.u.) is used throughout. The four Coulomb interaction terms: the first one is the nucleus-nucleus (NN) interaction, the fourth one is the electron-electron ($e_1 e_2$) interaction, the second one is the interaction between nucleus A and electron 2 (Ae), and the third one is the interaction between nucleus B and electron 1 (Be). Here $\vec{R}$ is the inter-nucleus displacement, $\vec{r}_1$ and $\vec{r}_2$ are the position vectors of the two system electrons with respect to their corresponding atomic nuclei.

The explicit form of the first term in the direct ($F_B^{NN}$) and rearrangement ($F_O^{NN}$) channels are:

$$F_B^{NN} = -\frac{\mu}{2\pi} \int d\vec{R} d\vec{r}_1 d\vec{r}_2 e^{-i\vec{k}_f \cdot \vec{R}'} \phi_{1s}^{A*}(\vec{r}_1) \phi_{1s}^{B*}(\vec{r}_2) \frac{Z_A Z_B}{R} e^{i\vec{k}_i \cdot \vec{R}} \phi_{1s}^A(\vec{r}_1) \phi_{1s}^B(\vec{r}_2) \qquad (2.5a)$$

$$F_O^{NN} = -\frac{\mu}{2\pi} \int d\vec{R} d\vec{r}_1 d\vec{r}_2 e^{-i\vec{k}_f \cdot \vec{R}_f} \phi_{1s}^{A*}(\vec{R}-\vec{r}_2) \phi_{1s}^{B*}(\vec{R}+\vec{r}_1) \frac{Z_A Z_B}{R} e^{i\vec{k}_i \cdot \vec{R}} \phi_{1s}^A(\vec{r}_1) \phi_{1s}^B(\vec{r}_2) \qquad (2.5b)$$

Similarly the fourth electron-electron correlation terms are,

$$F_B^{e_1 e_2} = -\frac{\mu}{2\pi} \int d\vec{R} d\vec{r}_1 d\vec{r}_2 e^{-i\vec{k}_f \cdot \vec{R}'} \phi_{1s}^{A*}(\vec{r}_1) \phi_{1s}^{B*}(\vec{r}_2) \frac{1}{|\vec{R}+\vec{r}_1-\vec{r}_2|} e^{i\vec{k}_i \cdot \vec{R}} \phi_{1s}^A(\vec{r}_1) \phi_{1s}^B(\vec{r}_2) \qquad (2.6a)$$

$$F_O^{e_1 e_2} = -\frac{\mu}{2\pi} \int d\vec{R} d\vec{r}_1 d\vec{r}_2 e^{-i\vec{k}_f \cdot \vec{R}_f} \phi_{1s}^{A*}(\vec{R}-\vec{r}_2) \phi_{1s}^{B*}(\vec{R}+\vec{r}_1) \frac{1}{|\vec{R}+\vec{r}_1-\vec{r}_2|} e^{i\vec{k}_i \cdot \vec{R}} \phi_{1s}^A(\vec{r}_1) \phi_{1s}^B(\vec{r}_2) \qquad (2.6b)$$

The second and third terms in direct and exchange channels are:

$$F_B^{Ae} = -\frac{\mu}{2\pi} \int d\vec{R} d\vec{r}_1 d\vec{r}_2 e^{-i\vec{k}_f \cdot \vec{R}'} \phi_{1s}^{A*}(\vec{r}_1) \phi_{1s}^{B*}(\vec{r}_2) \frac{(-Z_A)}{|\vec{R}-\vec{r}_2|} e^{i\vec{k}_i \cdot \vec{R}} \phi_{1s}^A(\vec{r}_1) \phi_{1s}^B(\vec{r}_2) \qquad (2.7a)$$

$$F_O^{Ae} = -\frac{\mu}{2\pi} \int d\vec{R} d\vec{r}_1 d\vec{r}_2 e^{-i\vec{k}_f \cdot \vec{R}_f} \phi_{1s}^{A*}(\vec{R}-\vec{r}_2) \phi_{1s}^{B*}(\vec{R}+\vec{r}_1) \frac{(-Z_A)}{|\vec{r}_1|} e^{i\vec{k}_i \cdot \vec{R}} \phi_{1s}^A(\vec{r}_1) \phi_{1s}^B(\vec{r}_2) \qquad (2.7b)$$

$$F_B^{Be} = -\frac{\mu}{2\pi} \int d\vec{R} d\vec{r}_1 d\vec{r}_2 e^{-i\vec{k}_f \cdot \vec{R}'} \phi_{1s}^{A*}(\vec{r}_1) \phi_{1s}^{B*}(\vec{r}_2) \frac{(-Z_B)}{|\vec{R}+\vec{r}_1|} e^{i\vec{k}_i \cdot \vec{R}} \phi_{1s}^A(\vec{r}_1) \phi_{1s}^B(\vec{r}_2) \qquad (2.8a)$$

$$F_O^{Be} = \frac{\mu}{2\pi} \int d\vec{R} d\vec{r}_1 d\vec{r}_2 e^{-i\vec{k}_f \cdot \vec{R}_f} \phi_{1s}^{A*}(\vec{R}-\vec{r}_2) \phi_{1s}^{B*}(\vec{R}+\vec{r}_1) \frac{(-Z_B)}{|\vec{r}_2|} e^{i\vec{k}_i \cdot \vec{R}} \phi_{1s}^A(\vec{r}_1) \phi_{1s}^B(\vec{r}_2) \qquad (2.8b)$$

with
$$\vec{R}' = \vec{R} + \frac{m_e}{m_A + m_e} \vec{r}_1 - \frac{m_e}{m_B + m_e} \vec{r}_2 \qquad (2.9a)$$

$$\vec{R}_f = \vec{R} + \frac{m_e}{m_A + m_e}(\vec{r}_2 - \vec{R}) - \frac{m_e}{m_B + m_e}(\vec{r}_1 + \vec{R}) \qquad (2.9b)$$

The notations $m_A$, $m_B$, $m_e$ represent the masses of the nucleus A, nucleus B and the mass of electron. The most difficult integral in eqn. (2.6b) contains three completely different coupled terms and two uncoupled terms in a nine-dimensional integral.

The effective range theory expresses the s-wave elastic phase shift ($\delta_0$) as a function of scattering-length ($a$) and projectile energy ($\sim k^2$) so that

$$k \cot \delta_0 = -\frac{1}{a} + \frac{1}{2} r_0 k^2 + O(k^4) \qquad (2.10)$$

when $k$ is the magnitude of the incident momentum and $r_0$ is the range of the potential.

### 3. Methodology

One can simplify all the above nine-dimensional integrals into the tractable two dimensional forms using the Bethe integrals and Fourier transforms [10]. A highly efficient computer code is developed using the FORTRAN programming language and numerical analysis. The code calculates the elastic phase-shifts and corresponding cross sections for both the singlet (+) and the triplet (-) channels. Two sets of coupled one-dimensional integral equation one $f_L^+$ corresponding to symmetric space part and the other $f_L^-$ corresponding to antisymmetric space part are solved parallelly in the same code using the matrix inversion method for each partial wave (L). The pole term in the coupled integral equation (2.2) is evaluated using the formulation with delta function and principal value parts so that

$$\frac{1}{\vec{k}_n^2 - \vec{k}''^2 + i\varepsilon} = -i\pi\delta(\vec{k}_n^2 - \vec{k}''^2) + \frac{P}{\vec{k}_n^2 - \vec{k}''^2} \qquad (3.1).$$

The principal value integral from zero to infinity has been replaced by

$$\int_0^\infty dk'' = \int_0^{2k_n} dk'' + \int_{2k_n}^\infty dk'' \qquad (3.2).$$

Even number of Gaussian points in the interval $0 - 2k_n$ are used to avoid the singularity problem at $k'' = k_n$. The Gauss-Legendre quadratures are used to perform the two-dimensional integrations in $B^\pm$ and the $\theta$ integration numerically in the present code. All the integrations converged properly when the magnitude of $k \leq 10$. The convergences are studied very carefully to integrate over the scattering angle ($\theta$) and the variables in the two-dimensional integrals (X and Y), varying the number of Gauss-Legendre quadratures. To study the correctness of the partial wave analysis method, the same partial wave analysis is applied to calculate the BO amplitude. It is verified that the BO amplitude obtained by partial wave analysis method and by the non-partial wave method are equal. More partial waves are required to get the converged data at higher energies. To include the contribution of the higher partial waves as accurately as possible in the integrated cross section, an augmented Born approximation [1-7,10-13] is used in the present code. To apply the augmented-Born method, we need to compare the SEM amplitude with the corresponding BO amplitude for each partial wave (L). As the value of L increases the difference between the two decreases gradually. When both are almost equal, only then it can replace the higher partial wave contribution of SEM amplitude by the equivalent BO amplitude. It is possible to get reliable data upto the value of $k \leq 10$ a.u. i.e. the energy E = 1360 eV in Ps-Ps scattering and E = 1.48 eV in H-H scattering following the coupled-channel methodology.

4. **Results and discussion**

We study the s-wave elastic phase-shift and the corresponding cross section (e.g. the amplitude square) for H-H scattering when both the atoms are in ground states. It is to be noted that both the partial wave phase-shifts and the partial wave cross sections are computed directly from the real and imaginary parts of the scattering amplitudes for the singlet ($f_L^+$) and the triplet ($f_L^-$) channels. A very large number of energy mesh-points are used in the energy interval $E = 1 \times 10^{-10} eV$ to $E = 1 \times 10^{-1} eV$ in search for Feshbach resonances. We find many resonances in the singlet channel and only one resonance in the triplet channel in the entire region. We have presented our data in the region $E = 1.4 \times 10^{-4}\, eV$ to $E = 1.2 \times 10^{-2}\, eV$ for both the singlet and triplet channels. In **Figures 1** and **2**, the s-wave elastic phase-shift and partial cross section data of H-H scattering using the present SEM code for both the singlet (S=0) and triplet (S=1) spin configurations of the system electrons are plotted against the incident momenta $k = 0.1$ to $k = 0.9$ a.u. The incident energy is related to $k$ by the relation $E(eV) = (27.21k^2/2\mu)$; $\mu = 918.5$ a.u. for H-H system. The Feshbach resonance in the triplet (-) channel is observed at $k \sim 0.55$ a.u. corresponding to the incident energy $E \sim 4.5 \times 10^{-3} eV$ and many resonances are found in the singlet channel. The tabular data for the s-, p- and d- wave phase-shifts are presented in **Table 1a.**

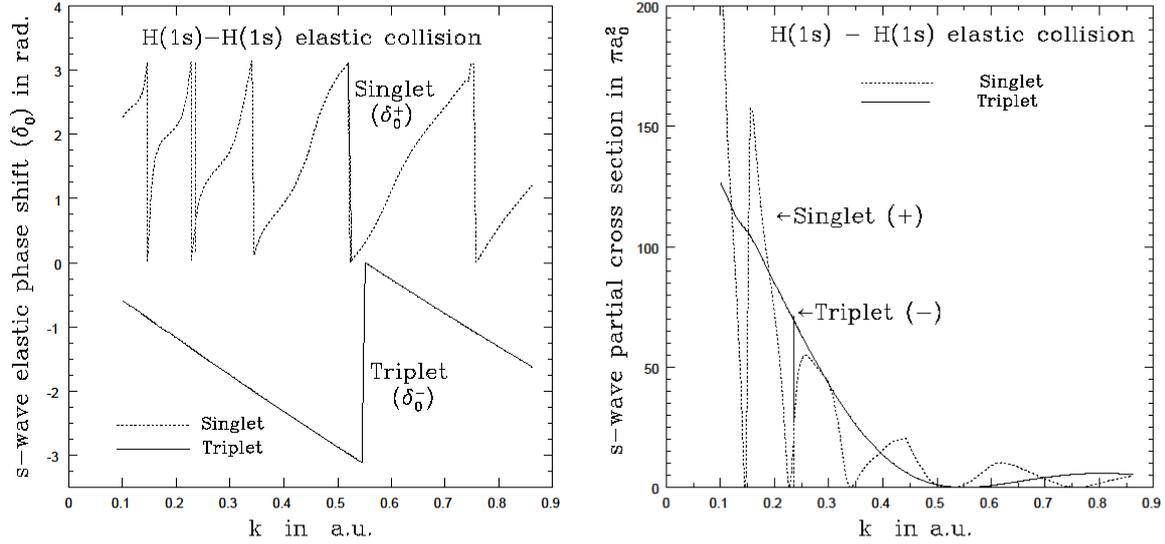

**Figure 1.** (a)The s-wave elastic phase shifts in radian and (b)The corresponding partial cross sections for the singlet (+) and triplet (-) channels against the values of incident momentum k in a.u. for H(1s)-H(1s) elastic collision.

**Table 1a.** The s-, p- and d- wave phase shifts in radian for both singlet (+) and triplet (-) channels in H(1s) and H(1s) elastic scattering.

| k (a.u.) | Singlet phase-shift in radian H and H system | | | Triplet phase-shift in radian H and H system | | |
|---|---|---|---|---|---|---|
| | s-wave | p-wave | d-wave | s-wave | p-wave | d-wave |
| 0.1 | 2.265 | 3.119 | 3.136 | -0.598 | -0.056 | -0.002 |
| 0.2 | 2.122 | 2.441 | 0.001 | -1.167 | -0.312 | -0.036 |
| 0.3 | 1.732 | 1.988 | 0.033 | -1.743 | -0.703 | -0.174 |
| 0.4 | 0.916 | 0.876 | 2.870 | -2.310 | -1.158 | -0.430 |
| 0.5 | 2.896 | 2.325 | 2.118 | -2.866 | -1.640 | -0.774 |
| 0.6 | 1.118 | 0.922 | 0.485 | -0.265 | -2.134 | -1.167 |
| 0.7 | 2.421 | 2.326 | 1.768 | -0.794 | -2.628 | -1.587 |
| 0.8 | 0.551 | 0.303 | 0.089 | -1.311 | -3.117 | -2.022 |
| 0.9 | 1.567 | 1.467 | 1.115 | -1.814 | -0.459 | -2.461 |

**Table 1b**. The s-, p- and d- wave cross sections in $\pi a_0^2$ for both singlet (+) and triplet (-) channels in H(1s) and H(1s) elastic scattering.

| k (a.u.) | Singlet (+) cross-section in $\pi a_0^2$ H and H system | | | Triplet (-) cross-section in $\pi a_0^2$ H and H system | | |
|---|---|---|---|---|---|---|
| | s-wave | p-wave | d-wave | s-wave | p-wave | d-wave |
| 0.1 | 236.131 | 0.619 | 0.050 | 126.716 | 3.759 | 0.005 |
| 0.2 | 72.529 | 124.678 | 0.001 | 84.581 | 28.319 | 0.649 |
| 0.3 | 43.301 | 111.466 | 0.248 | 43.133 | 55.662 | 6.632 |
| 0.4 | 15.727 | 44.249 | 8.973 | 13.650 | 62.931 | 21.768 |
| 0.5 | 0.942 | 25.509 | 58.335 | 1.187 | 47.767 | 39.068 |
| 0.6 | 8.987 | 21.15 | 12.091 | 0.765 | 23.840 | 46.988 |
| 0.7 | 3.550 | 12.973 | 39.252 | 4.152 | 5.915 | 40.805 |
| 0.8 | 1.712 | 1.669 | 0.247 | 5.837 | 0.011 | 25.311 |
| 0.9 | 4.938 | 14.656 | 19.907 | 4.650 | 2.904 | 9.766 |

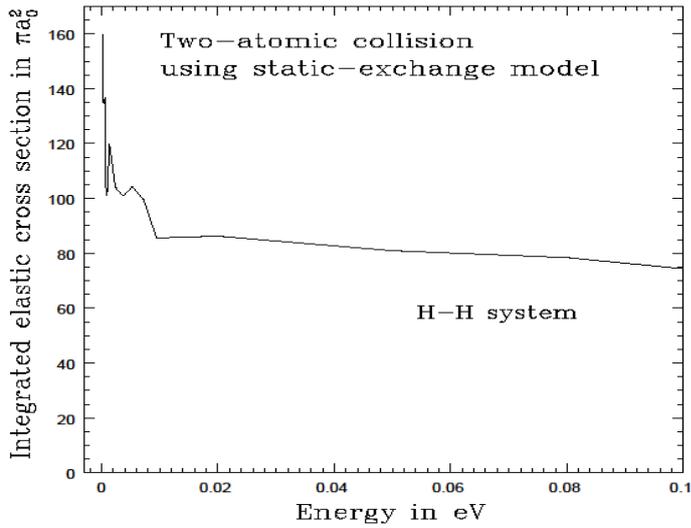

**Figure 2.** The integrated/total elastic cross sections in $\pi a_0^2$ of H(1s)-H(1s) elastic scattering.

The s-, p- and d- wave partial cross sections are presented in tabular form in the **Table 1b**. All the three partial cross sections for L= 0, 1 and 2 are contributing appreciably to the integrated elastic cross section plotted in **Figure 2**; but at k = 0.1 a.u. the s-wave have the maximum contribution to the total / integrated elastic cross section.

In Ps(1s)–Ps(1s) collision, the s-wave elastic phase shifts of both the total spin aligned states: S=0 for the singlet (+) and S=2 for the triplet (-) are presented in **Figure 3** against the incident momenta $k = 0.1$ to $k = 0.6$ a.u. at the energy region below the threshold. These data are compared with the data of Ivanov et al [14] in **Figure 3**. In **Table 2a**, the s-, p- and d- wave elastic phase shifts and in **Table 2b**, the corresponding s-, p- and d- wave partial cross sections are presented following the present exact-exchange analysis in two Ps collision. It should be noted that all the p-wave (odd parity) phase shifts are zero or $\pm \pi$ radian and all the p-wave cross sections are zero. In the present new code when the most difficult electron-electron interaction matrix-element is approximated by the positron-positron interaction matrix-element, the s- (even parity) and d- (even parity) wave data are exactly the same with the exact-exchange analysis data of **Table 2a and 2b**, but all the p-wave (odd parity) data are very different and partial cross sections are not zero or close to zero. The present s-, p- and d- wave data using the approximate form for the electron-electron exchange term are presented in **Tables 3a and 3b**. Comparing the phase shifts data of **Tables 2a & 3a** and partial cross sections data of **Tables 2b and 3b**, it is evident that the approximation to define the electron-electron exchange term introduced a large error to calculate the contribution of highly important p-wave (odd parity) e.g. the non-spherical orbitals. In addition, a very large number of energy mesh-points in the energy inter-val 0.13 to 5.10 eV are used to calculate the s-wave elastic phase-shifts and the corresponding partial cross sections in search of Feshbach resonances using the exact-analysis of exchange, but no resonances are found. In **Figure 4** the integrated/ total elastic cross section $\sigma$ using the augmented-Born approximation and the quenching cross section $\sigma_q$ for the Ps(1s)-Ps(1s) elastic scattering are presented in the energy region 0 to 150 eV. A resonance like structure - a deep and a peak appears in the energy region 20 - 30 eV in both the total elastic and quenching cross sections in **Figure 4.** In **Figure 5** the ortho to para conversion ratios ($\sigma/\sigma_q$) are plotted against the incident energy and compared with the Ps-H system. It is evident from the figure that most of the annihilations occur at lower energies below 50 eV.

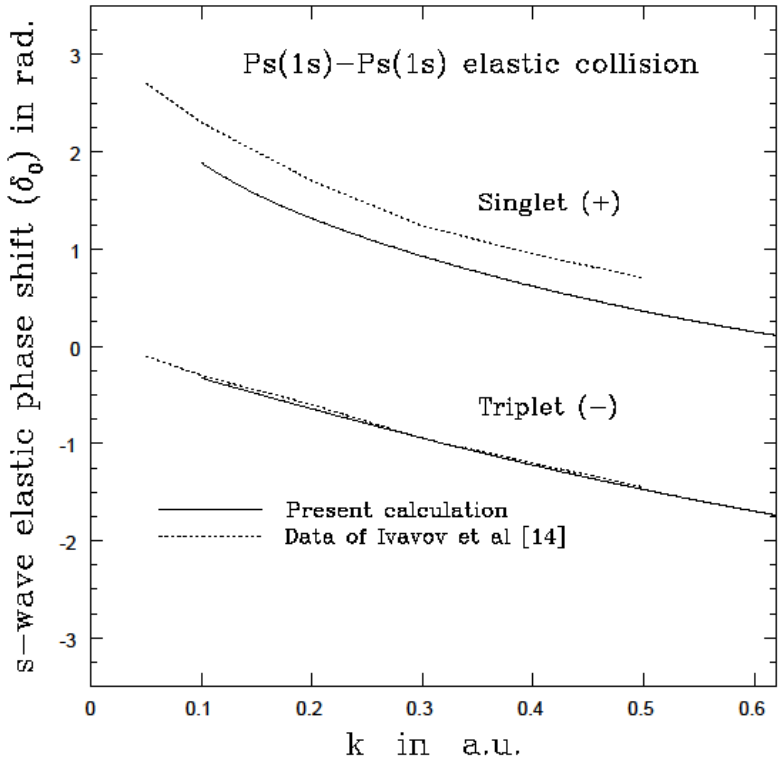

**Figure 3.** The s-wave elastic phase shifts in radian for both singlet (+) and triplet (-) channels in Ps(1s)-Ps(1s) scattering against the incident momentum k in a.u.

**Table 2a.** The s-, p- and d- wave phase shifts in radian for both singlet (+) and triplet (-) channels in Ps(1s) and Ps(1s) elastic scattering.

| k (a.u.) | Singlet (+) phase-shift in radian Ps(1s) and Ps(1s) system | | | Triplet (-) phase-shift in radian Ps(1s) and Ps(1s) system | | |
|---|---|---|---|---|---|---|
| | s-wave | p-wave | d-wave | s-wave | p-wave | d-wave |
| 0.1 | 1.887 | 3.142 | 3.141 | -0.325 | -3.142 | -3.141 |
| 0.2 | 1.313 | 3.142 | 3.141 | -0.643 | -3.142 | -3.141 |
| 0.3 | 0.926 | 3.142 | 3.138 | -0.944 | -3.142 | -3.137 |
| 0.4 | 0.616 | 3.142 | 3.129 | -1.222 | -3.142 | -3.127 |
| 0.5 | 0.359 | 3.142 | 3.114 | -1.474 | -3.142 | -3.109 |
| 0.6 | 0.147 | 3.142 | 3.095 | -1.697 | -3.142 | -3.085 |
| 0.7 | 3.118 | 3.142 | 3.073 | -1.892 | -3.142 | -3.059 |
| 0.8 | 2.986 | 3.142 | 3.052 | -2.059 | -3.142 | -3.035 |
| 0.9 | 2.890 | 3.142 | 3.035 | -2.200 | -3.142 | -3.017 |

**Table 2b.** The s-, p- and d- wave cross sections in $\pi a_0^2$ for both singlet (+) and triplet (-) channels in Ps(1s) and Ps(1s) elastic scattering.

| k (a.u.) | Singlet (+) cross-section in $\pi a_0^2$ Ps(1s) - Ps(1s) system | | | Triplet (-) cross-section in $\pi a_0^2$ Ps(1s) - Ps(1s) system | | |
|---|---|---|---|---|---|---|
| | s-wave | p-wave | d-wave | s-wave | p-wave | d-wave |
| 0.1 | 361.227 | 0.000 | 0.0000 | 41.016 | 0.000 | 0.0000 |
| 0.2 | 93.524 | 0.000 | 0.0002 | 35.966 | 0.000 | 0.0002 |
| 0.3 | 28.383 | 0.000 | 0.003 | 29.147 | 0.000 | 0.004 |
| 0.4 | 8.350 | 0.000 | 0.019 | 22.081 | 0.000 | 0.026 |
| 0.5 | 1.978 | 0.000 | 0.059 | 15.850 | 0.000 | 0.083 |
| 0.6 | 0.240 | 0.000 | 0.122 | 10.934 | 0.000 | 0.177 |
| 0.7 | 0.044 | 0.000 | 0.191 | 7.349 | 0.000 | 0.278 |
| 0.8 | 0.149 | 0.000 | 0.248 | 4.874 | 0.000 | 0.351 |
| 0.9 | 0.306 | 0.000 | 0.281 | 3.229 | 0.000 | 0.378 |

**Table 3a.** The s-, p- and d- wave phase shifts in radian for both singlet (+) and triplet (-) channels in Ps(1s) and Ps(1s) elastic scattering approximating the electron-electron interaction term.

| k (a.u.) | Singlet (+) phase-shift in radian Ps(1s) and Ps(1s) system | | | Triplet (-) phase-shift in radian Ps(1s) and Ps(1s) system | | |
|---|---|---|---|---|---|---|
| | s-wave | p-wave | d-wave | s-wave | p-wave | d-wave |
| 0.1 | 1.887 | 0.0153 | 3.141 | -0.326 | -0.0064 | -3.141 |
| 0.2 | 1.313 | 0.1210 | 3.141 | -0.643 | -0.0433 | -3.141 |
| 0.3 | 0.926 | 0.3564 | 3.138 | -0.944 | -0.1154 | -3.137 |
| 0.4 | 0.616 | 0.5967 | 3.129 | -1.222 | -0.2086 | -3.127 |
| 0.5 | 0.359 | 0.7125 | 3.114 | -1.474 | -0.3053 | -3.109 |
| 0.6 | 0.147 | 0.7245 | 3.095 | -1.697 | -0.3917 | -3.085 |
| 0.7 | 3.118 | 0.6843 | 3.073 | -1.892 | -0.4587 | -3.059 |
| 0.8 | 2.986 | 0.6233 | 3.052 | -2.059 | -0.5019 | -3.035 |
| 0.9 | 2.890 | 0.5568 | 3.035 | -2.200 | -0.5198 | -3.017 |

**Table 3b.** The s-, p- and d- wave cross sections in $\pi a_0^2$ for both singlet (+) and triplet (-) channels in Ps(1s) and Ps(1s) elastic scattering approximating the electron-electron interaction term.

| k (a.u.) | Singlet (+) cross-section in $\pi a_0^2$ Ps(1s) - Ps(1s) system | | | Triplet (-) cross-section in $\pi a_0^2$ Ps(1s) - Ps(1s) system | | |
|---|---|---|---|---|---|---|
| | s-wave | p-wave | d-wave | s-wave | p-wave | d-wave |
| 0.1 | 361.227 | 0.282 | 0.0000 | 41.016 | 0.0492 | 0.0000 |
| 0.2 | 93.524 | 4.368 | 0.0002 | 35.966 | 0.5620 | 0.0002 |
| 0.3 | 28.383 | 16.232 | 0.003 | 29.147 | 1.7692 | 0.004 |
| 0.4 | 8.350 | 23.685 | 0.019 | 22.081 | 3.2178 | 0.026 |
| 0.5 | 1.978 | 20.513 | 0.059 | 15.850 | 4.3372 | 0.083 |
| 0.6 | 0.240 | 14.641 | 0.122 | 10.934 | 4.8572 | 0.177 |
| 0.7 | 0.044 | 9.787 | 0.191 | 7.349 | 4.8024 | 0.278 |
| 0.8 | 0.149 | 6.389 | 0.248 | 4.874 | 4.3393 | 0.351 |
| 0.9 | 0.306 | 4.137 | 0.281 | 3.229 | 3.6558 | 0.378 |

In **Table 4**, the variation of scattering length with reduced-mass for different Ps-atom systems (e.g. Ps-Ps, Ps-Mu, Ps-H, Ps-D, Ps-T) are presented for both the singlet and triplet channels. A systematic variation of scattering length with the reduced mass of the system is observed. This observed dependence of scattering length with reduced-mass could explain the electron-like scattering of Ps [15] since the reduced-mass of Ps-atom and electron-atom systems are almost the same and the difference decreases as the atom becomes heavier. The variation of triplet scattering length with reduced-mass for the heavier atomic systems e.g. Mu-Mu, Mu-H, Mu-D, Mu-T, H-H, H-D, H-T, D-D, D-T, T-T are presented in **Table 5**. Again we are finding a systematic dependence of scattering length with the reduced-mass of the system [5]. It should be noted that the SEM theory include only the non-adiabatic short-range effects due to exchange, but no long-range interaction. **Figure 6** shows the plot of the effective range theory used to calculate the scattering lengths for heavier systems.

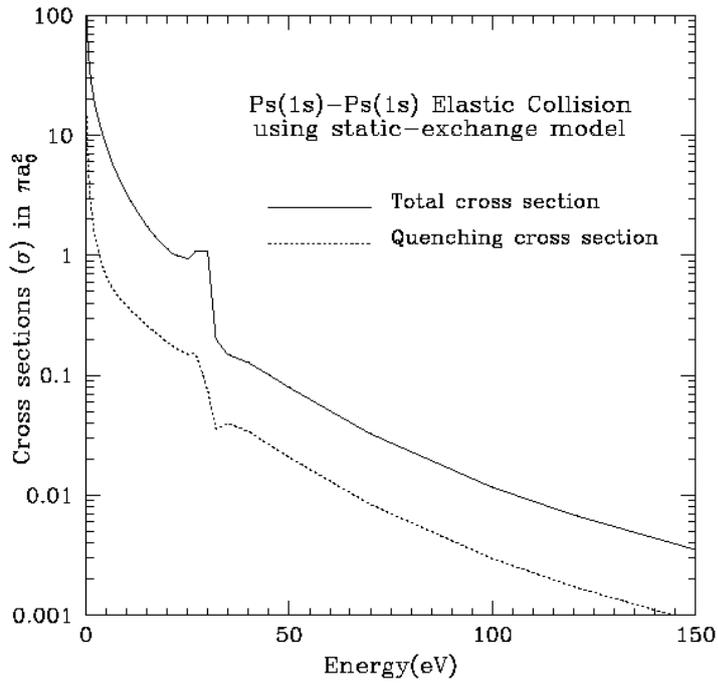

**Figure 4.** The variation of integrated /total elastic cross sections and the quenching cross section in $\pi a_0^2$ of Ps(1s)-Ps(1s) scattering with energy in eV.

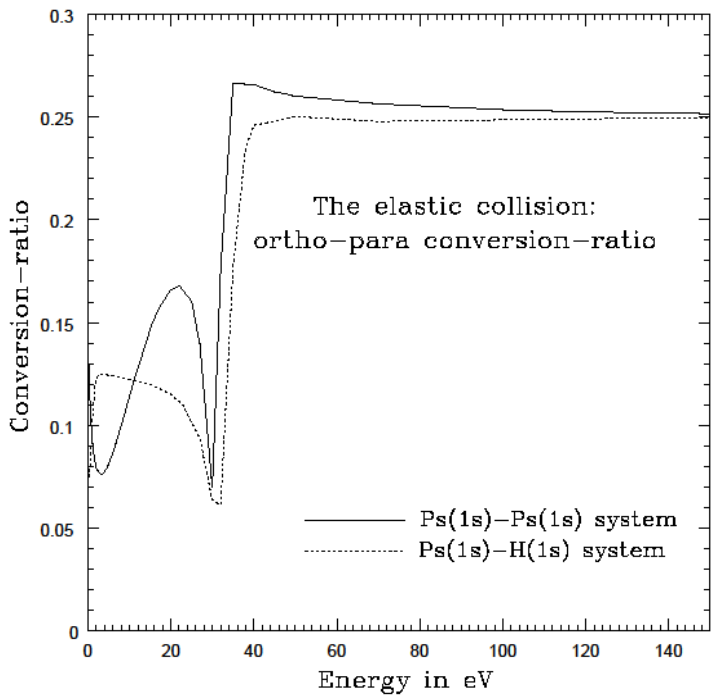

**Figure 5.** The comparison of the variation of ortho to para conversion ratios ($\sigma_q/\sigma$) of Ps(1s)-Ps(1s) and Ps(1s)-H(1s) systems with energy.

**Table 4.** The variation of scattering length with reduced mass of the system.

| Systems | | Ps-Ps | Ps - Mu | Ps - H | Ps - D | Ps - T |
|---|---|---|---|---|---|---|
| Reduced Masses ( a. u.) | | 1.0 | 1.9809 | 1.9978 | 1.9989 | 1.9993 |
| Scattering lengths (a.u.) | Singlet (+) | 9.35 | 7.40 | 7.24 | 7.18 | 7.14 |
| | Triplet (-) | 3.25 | 2.50 | 2.48 | 2.46 | 2.45 |

**Table 5.** The variation of triplet scattering length with reduced mass of different H-like systems.

| System → | Mu-Mu | Mu-H | Mu-D | Mu-T | H-H | H-D | H-T | D-D | D-T | T-T |
|---|---|---|---|---|---|---|---|---|---|---|
| Reduced mass → in a.u. | 103.9 | 186.7 | 196.7 | 200.2 | 918.5 | 1224.5 | 1377.6 | 1836.5 | 2203.7 | 2754.5 |
| Scattering length → in a.u. | 4.54 | 4.76 | 4.88 | 4.95 | 5.88 | 6.25 | 6.37 | 6.58 | 6.68 | 6.90 |

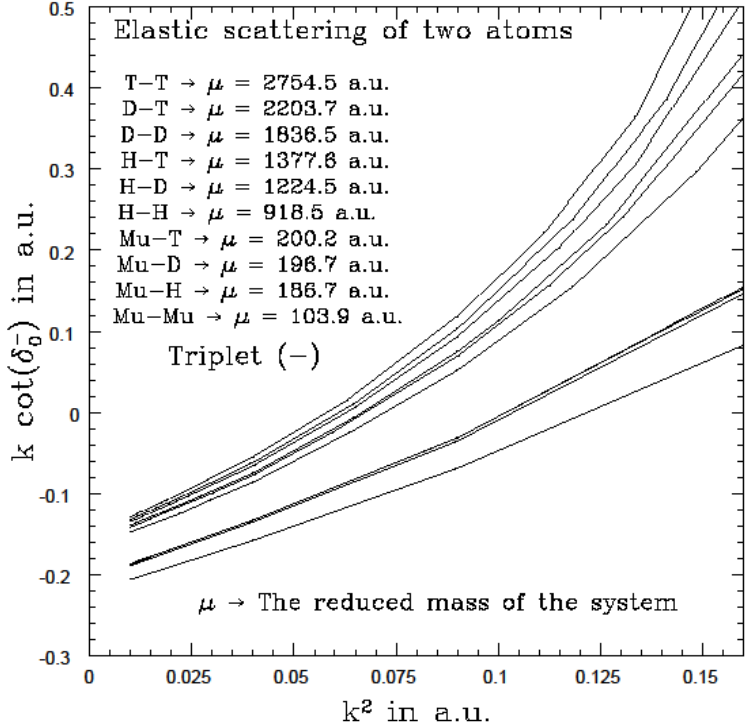

**Figure 6.** $k \cot \delta_0^-$ vs $k^2$ curve for different H-like systems. The reduced masses are increasing gradually for the systems from bottom to top solid curves.

**Table 6.** The scattering length in atomic units using SEM and MSEM for different values of $R_0$.

| Scattering length in atomic unit (a.u.) | | | | | | | | | | | | | |
|---|---|---|---|---|---|---|---|---|---|---|---|---|---|
| Using SEM | Using MSEM with $R_0 =$ | | | | | | | | | | | | Data of others |
| | $20a_0$ | $15a_0$ | $12a_0$ | $11a_0$ | $10a_0$ | $9a_0$ | $8a_0$ | $7a_0$ | $6a_0$ | $5a_0$ | $4a_0$ | $3a_0$ | $2a_0$ | |
| 5.88, 5.90[a] | 5.80 | 5.68 | 5.26 | 5.11 | 4.89 | 4.63 | 4.38 | 4.03 | 3.77 | 3.68 | 3.63 | 3.60 | 3.58 | 2.04[a], 1.91[b], 1.22[c], 1.34[d], 1.3[e] |

[a]Sen, Chakraborty and Ghosh [17];
[b]Jamieson, Dalgarno and Yukich [18];
[c]Jamieson and Dalgarno [19];
[d]Williams and Julienne [20];
[e]Koyama and Baird [21].

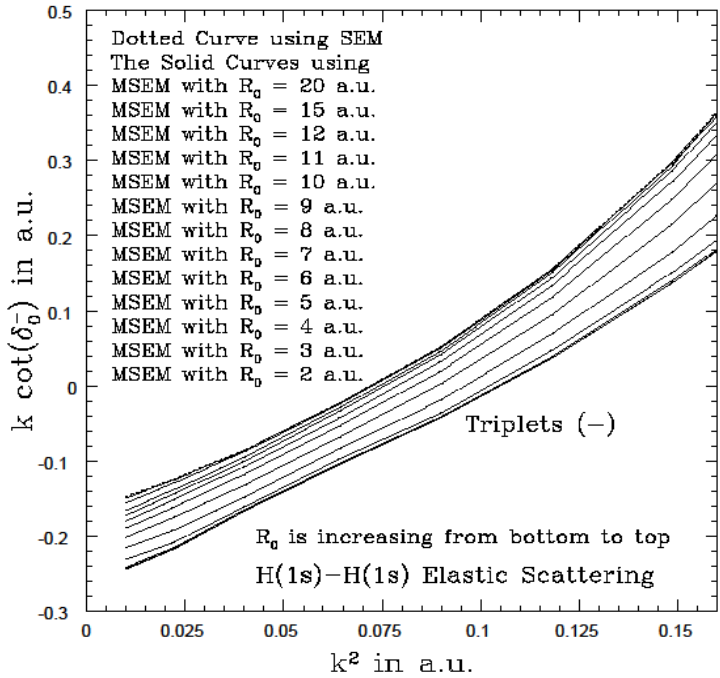

**Figure 7.** $k \cot \delta_0^-$ vs $k^2$ plot for H(1s)-H(1s) elastic scattering using SEM and MSEM for $R_0 = 2a_0$, $3a_0$, $4a_0$, $5a_0$, $6a_0$, $7a_0$, $8a_0$, $9a_0$, $10a_0$, $11a_0$, $12a_0$, $15a_0$, $20a_0$.

The MSEM code is used to study the effect of long-range attractive van der Waals interaction in H(1s)-H(1s) elastic scattering. The most accurate value for the van der Waals coefficient [16] is used to define the long-range potential. Different minimum values of interatomic distance $R_0$ are chosen to vary the strength of interatomic potential. The variation of scattering length with the variation of values of $R_0$ are tabulated and compared with SEM data in **Table 6**. With the increase of the values of $R_0$, the data are being gradually closer to the SEM data. As the value of $R_0$ decreases, the strength of attractive van der Waals interaction gradually increases. It is to be observed that the scattering length is gradually decreasing and being closer to the reported accurate values [17-21] as the strength of van der Waals interaction increases with decreasing $R_0$. In **Figure 7**, the plot applying effective range theory for different values of $R_0$ is presented. The systematic variation of scattering length towards the accurate values with decreasing $R_0$ has a great significance to understand the basic physics of two-atomic systems.

All the findings are consistent with the basic physics that the stronger attractive potential causes shorter scattering length and the stronger repulsive potential causes longer scattering length [22]. Again the strength of an interaction potential increases with the increase of mass of the system [22]. So in heavier systems the attractive potential should be more stronger. In lighter systems like Ps-Mu, Ps-H, Ps-D, Ps-T or in Ps-Ps, the variation in interatomic potential due to the reduced mass of the system will be very small or negligible. At low and cold energies, the long-range van der Waals interaction has an important role in addition to short-range non-adiabatic effects due to exchange. So inclusion of both these effects are extremely useful to describe the cold energy atomic collision processes. The data using the appropriate values of the van der Waals coefficients for different two-atomic systems in MSEM code could explain the unexplained findings like the occurrence of Bose-Einstein condensation (BEC) [23] in $Rb_{85}$ - $Rb_{85}$ system with $\mu = 78048.5$ a.u. The reduced masses of $Rb_{87}$ - $Rb_{87}$ system is $\mu = 79884.5$ a.u. and for $Rb_{85}$ - $Rb_{87}$ system is $\mu = 78955.8$ a.u., both are greater than the reduced mass of $Rb_{85}$ - $Rb_{85}$ system. Again it is useful to note that only the positive scattering length is useful to find the BEC in a system.

It would be extremely useful to include the first excited 2s and 2p states of both the atoms in the coupled-channel methodology for more accurate data; indeed it is very difficult and tedious job. Again it would also be useful to introduce the first excited 2s state of both the atoms following the coupled-channel methodology in the MSEM code to include the majority of the non-adiabatic short-range effects. Again it is a good work to calculate the most accurate values of the van der Waals coefficients $C_6$ for different H-like two-atomic systems and for the alkali atomic systems, those are useful to improve the present work and to invent new physics.

## 5. Conclusion

In the present review, a thorough studies are made to summarize the recent findings using the exact analysis of the two H-like atomic collisions considering the system as a four-body Coulomb problem in the center of mass frame using the SEM and MSEM introduced recently by Ray [1-5]. The present knowledge is highly useful to understand the basic fundamental physics and to answer many unanswered questions [15,23]. One could improve the code including more channels according to necessity in the coupled-channel methodology adapted here to search new physics.

**Acknowledgement:**

The author acknowledges the support of DST, India through grant number SR/WOSA/PS-13/2009.